\documentclass[]{spie}

\pdfoutput=1 

\usepackage[]{graphicx,color}
\usepackage[]{amsmath}
\usepackage[]{amssymb}
\usepackage[]{deluxetable}

\def\gtrsim{\mathrel{\hbox{\rlap{\hbox{\lower4pt\hbox{$\sim$}}}\hbox{$>$}}}}

\title{ALES: Overview and Upgrades}

\author{
Andrew J. Skemer\authorinfo{\supit{*} Contact: askemer@ucsc.edu}\supit{a},
Philip Hinz\supit{b},
Jordan Stone\supit{b},
Michael Skrutskie\supit{c}
Charles E. Woodward\supit{d}
Jarron Leisenring\supit{b}
and
Zackery Briesemeister\supit{a}
\skiplinehalf
\supit{a} University of California, Santa Cruz, USA;
\skiplinehalf
\supit{b} University of Arizona, USA;
\skiplinehalf
\supit{b} University of Virginia, USA;
\skiplinehalf
\supit{b} Minnesota Institute for Astrophysics, University of Minnesota, USA;
}

\begin{document} 
\maketitle

\begin{abstract}
The Arizona Lenslets for Exoplanet Spectroscopy (ALES) is the world’s first AO-fed thermal infrared integral field spectrograph, mounted inside the Large Binocular Telescope Interferometer (LBTI) on the LBT.  An initial mode of ALES allows 3-4 $\mu$m spectra at R$\sim$20 with 0.026” spaxels over a 1”x1” field-of-view.  We are in the process of upgrading ALES with additional wavelength ranges, spectral resolutions, and plate scales allowing a broad suite of science that takes advantage of ALES’s unique ability to work at wavelengths $> $2 microns, and at the diffraction limit of the LBT’s full 23.8 meter aperture.
\end{abstract}
\keywords{Adaptive optics, integral field spectroscopy, exoplanet imaging, exoplanet instrumentation}

\section{INTRODUCTION}
Integral field spectrographs (IFS’s) have become ubiquitous on large adaptive optics (AO) telescopes, in particular for their ability to obtain spectra of directly-imaged exoplanets.  Most of these IFS's operate in the near-infrared (1-2$\mu$m).  ALES is the world's first adaptive optics IFS that operates in the thermal infrared (3-5$\mu$m), where gas-giant planets peak in brightness, and various molecular features of exoplanets, circumstellar disks and Solar System bodies become accessible.  

The original implementation of ALES\cite{2015SPIE.9605E..1DS} comprised a small field-of-view lenslet array, and a single plate-scale and disperser setting.  We are in the process of upgrading ALES to (1) increase its field-of-view, (2) add plate scales appropriate for seeing-limited, adaptive optics and interferometric observations, and (3) add new disperser modes with R$\sim$10-200 resolution across various bandpasses from 1.5-5 $\mu$m.  The original implementation of ALES is described in Section 2.  The upgraded ALES, which will be installed in Summer 2018, is described in Section 3.

\section{ORIGINAL IMPLEMENTATION OF ALES}
ALES was designed to fit into the LBTI\cite{2012SPIE.8445E..0UH,2014SPIE.9146E..0TH,2016SPIE.9907E..04H}/LMIRcam\cite{2010SPIE.7735E..3HS,2008SPIE.7013E..3AW,2012SPIE.8446E..4FL} light-path, taking advantage of an existing AO system\cite{2011SPIE.8149E..02E,2011SPIE.8149E..02E,2014SPIE.9148E..03B}, dewar, camera optics, filter wheels, and HAWAII-2RG detector, along with 3 pairs of intermediate focal and pupil planes (see Figure 1).  The original implementation of ALES has a $50 \times 50$ spaxel, 1.2 square arcsecond field of view, spanning 2.8-4.2$\mu$m, with R $\sim 20$ spectral resolution. Such a configuration is ideal for taking low-resolution spectra of exoplanets across broad molecular features, and also covers the 3.1 $\mu$m ice feature, 3.3 $\mu$m PAH feature, and 4.1 $\mu$m Br-alpha emission line.  The optics comprising ALES (listed below) are installed in filter wheels making ALES a completely modular design.

\begin{figure}[htbp]
\begin{center}
  \hbox{
    \hspace{0.1in}
      \includegraphics[angle=0,width=1.0\linewidth]{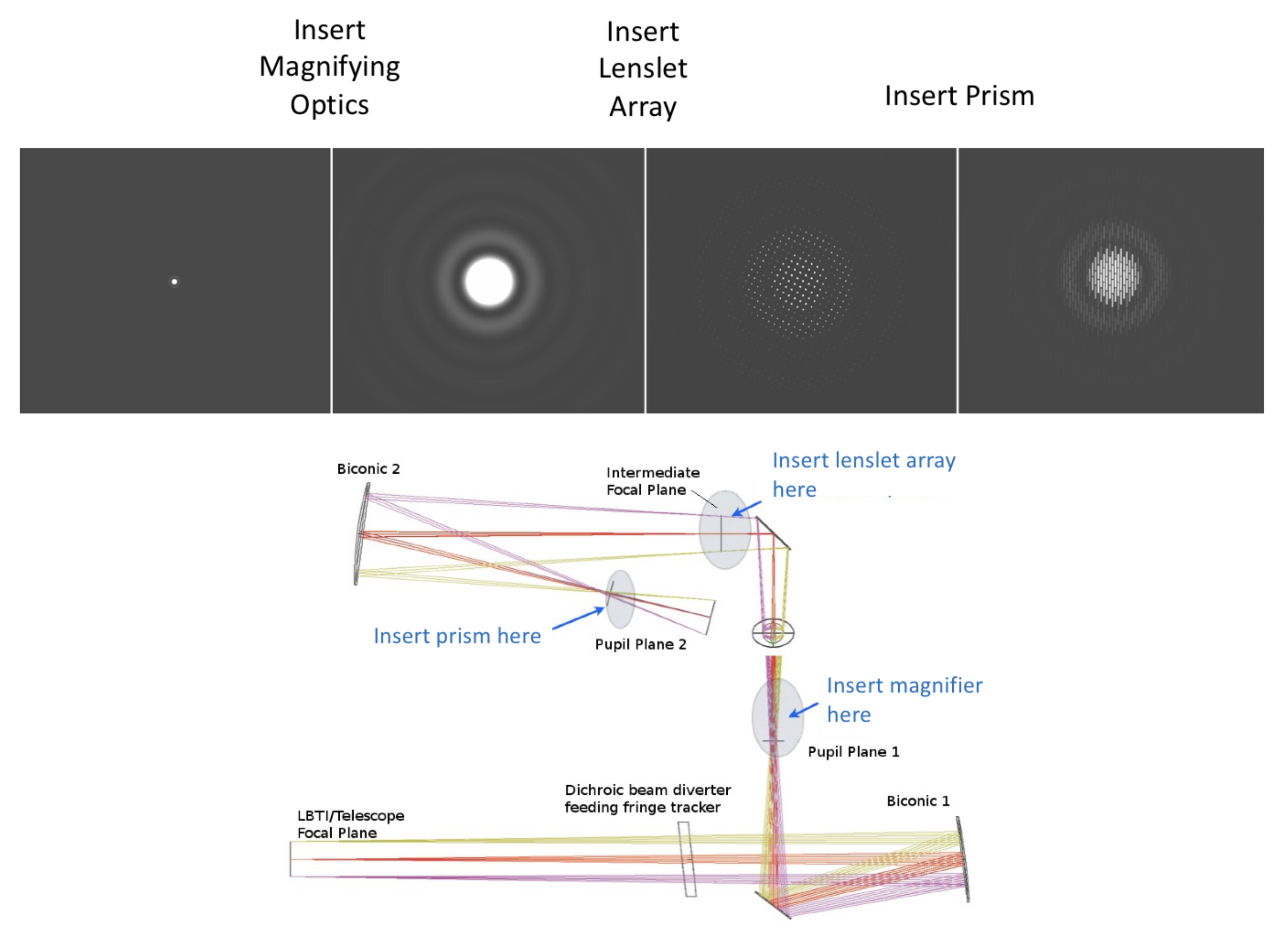}}
\end{center}
      \vspace{0.0in}
      \caption{--- Top: Simulation of ALES optics modifying the image at the LMIRcam focal plane as they are inserted.  Bottom: Location of the ALES optics in the LMIRcam optical design.
	  } \label{fig:ALES schematic}
\end{figure}

\begin{itemize}
\item \textbf{Distortion grid}—a stock part from Thorlabs consisting of chrome dots on a glass substrate separated by 250 $\mu$m.  This part is installed in an intermediate focal plane before the rest of the ALES optics to calibrate geometric distortions.
\item \textbf{Narrowband filters}—Cryogenically calibrated narrowband filters that can be used in series with the rest of the ALES optics to calibrate the wavelength solutions of the individual spectra.
\item \textbf{8x Keplerian magnifier}—A ZnSe/CaFl achromatic Keplerian magnifier to change LMIRcam’s native plate scale to one that is appropriate for the lenslet array.
\item \textbf{Lenslet array}—a 50x50 array of 360x360 $\mu$m lenslets on a silicon substrate that each produce f/15 beams (limited by LMIRcam’s downstream optics) imaged onto an intermediate focal plane.  To suppress diffractive crosstalk between the adjacent spectra, a pinhole grid is aligned and glued at a fixed separation from the lenslet array.
\item \textbf{Direct-vision prism}—a sapphire/ZnSe doublet prism that disperses each spaxel image into a 35 pixel spectrum on the HAWAII-2RG detector.  The dispersion direction is rotated 26.6 degrees with respect to the lenslet array to keep the spectra from overlapping.
\item \textbf{Blocking filter}—a 2.8-4.2 $\mu$m filter that limits the prism dispersion to a 35 pixel spectrum to keep the spectra from overlapping.
\end{itemize}

ALES had first light in June 2015.  Raw and extracted images of Io are shown in Figure 2 and 3, demonstrating ALES’s functionality\cite{2015SPIE.9605E..1DS}.  Additional science demonstration images of exoplanets and brown dwarf binaries are presented in Stone et al. (these proceedings)\cite{Stone_2018SPIE} and Briesemeister et al. (these proceedings)\cite{Briesemeister_2018SPIE}.

\begin{figure}[htbp]
\begin{center}
  \hbox{
    \hspace{0.1in}
      \includegraphics[angle=0,width=1.0\linewidth]{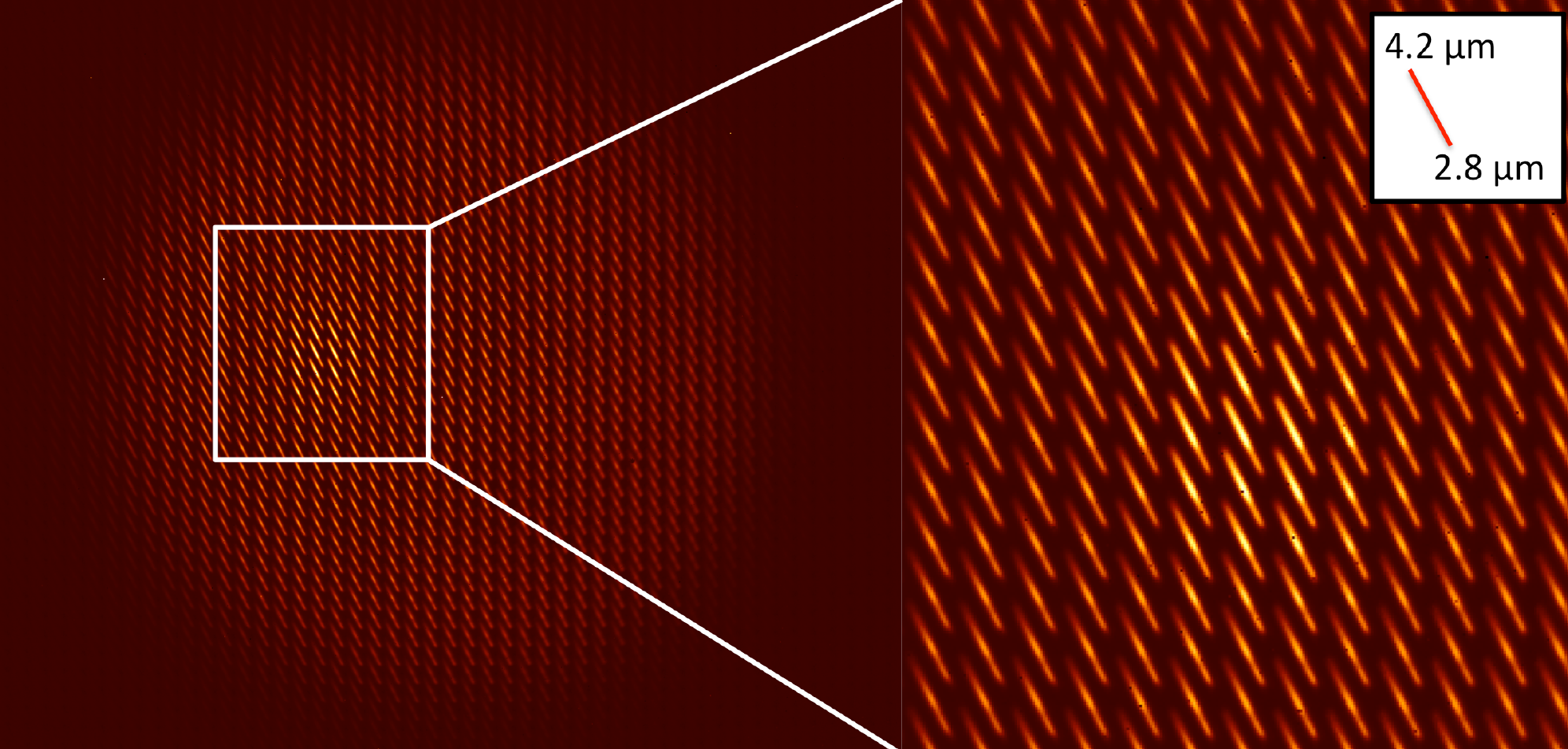}}
\end{center}
      \vspace{0.0in}
      \caption{--- Left:Unprocessed ALES image of Io. Right: Zoom-in on the Loki Patera volcano.
	  } \label{fig:Io 1}
\end{figure}

\begin{figure}[htbp]
\begin{center}
  \hbox{
    \hspace{0.1in}
      \includegraphics[angle=0,width=1.0\linewidth]{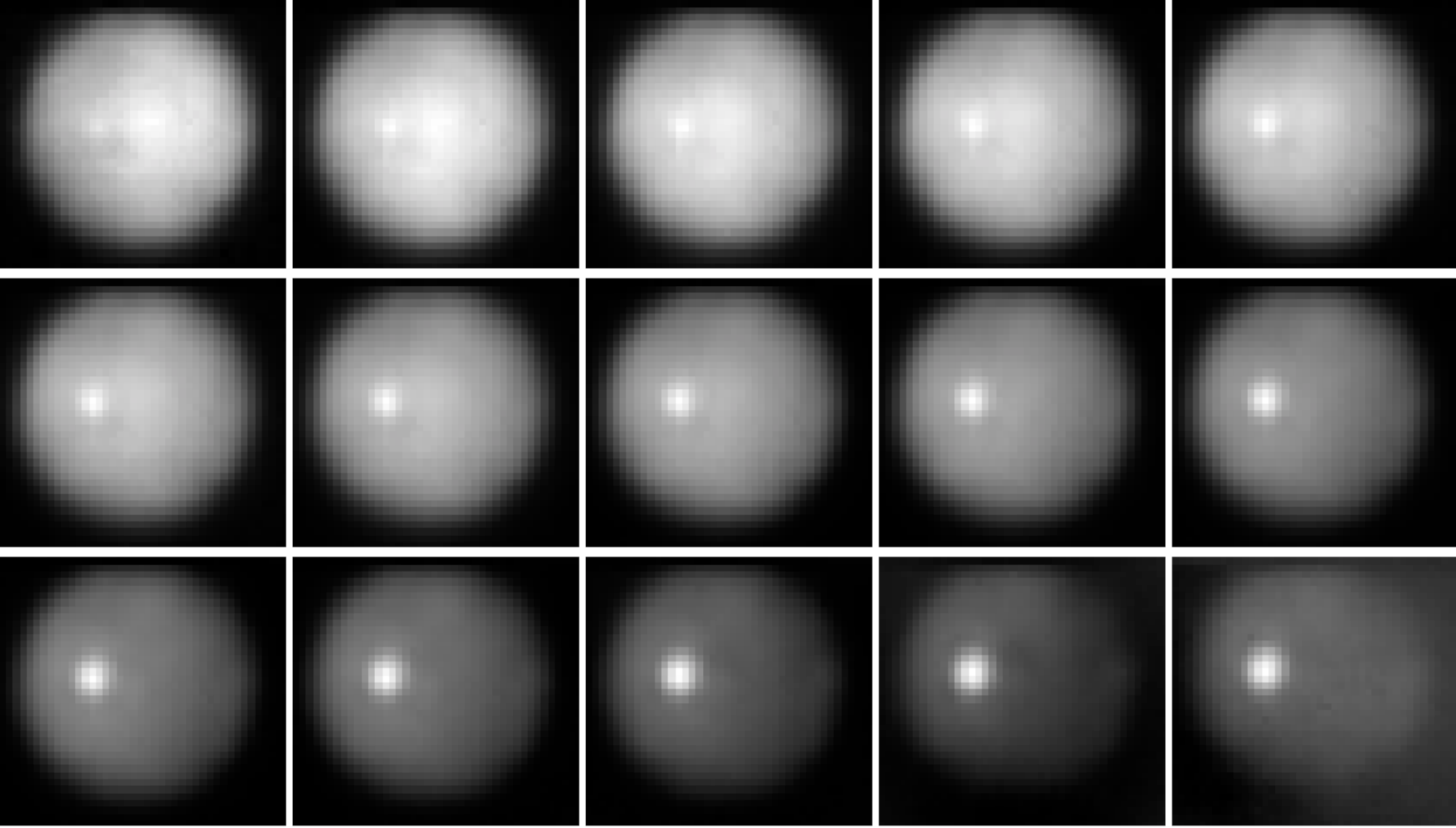}}
\end{center}
      \vspace{0.0in}
      \caption{--- Extracted ALES images of Io with wavelengths ranging from 2.8$\mu$m (upper left) to 4.2$\mu$m (lower right) by steps of 0.1$\mu$m. Light from Io’s disk is primarily reflected Sun-light. Thermal emission from the volcano peaks at longer wavelengths.
	  } \label{fig:Io 2}
\end{figure}

\section{UPGRADES AND UPGRADE STATUS}
Our motivations for upgrading ALES are to expand its capabilities and improve its performance.  The upgrades, described below, will increase ALES's field-of-view, increase its spectral resolution, allow selectable bandpasses, and allow selectable plate scales.  A complete description of the upgraded ALES optical design is presented in Hinz et al. (these proceedings)\cite{Hinz_2018SPIE}.

\subsection{DETECTOR ELECTRONICS}
LMIRcam has a HAWAII-2RG detector and optics that pass the full area imaged onto the 2048x2048 detector.  However, the heritage of LMIRcam is such that the electronics used with the original version of ALES only read out a 1024x1024 subregion of the array\cite{2012SPIE.8446E..4FL}.  We have upgraded the LMIRcam electronics with Teledyne's SIDECAR electronics, and now read the entire array.

\subsection{LENSLET ARRAY}
The ALES concept involves placing a lenslet array at the LMIRcam intermediate focal plane (see Figure 1).  This has allowed us to rapidly deploy a new capability within an existing instrument.  However, it has resulted in a non-optimal optical design: a single biconic reimaging the lenslet array focus onto the detector creates field dependent astigmatism that increases off-axis.  The upgraded version of ALES has a lenslet array that pre-corrects the field dependent astigmatism by defining a different aspheric lenslet description for each lenslet.  

The combination of the detector electronics upgrade and the ability of the new lenslet array to work off-axis allows us to greatly increase ALES's field-of-view.  There is a trade-off between field-of-view and spectral resolution in a lenslet integral-field spectrograph, and we therefore decided to use the additional area to both increase field-of-view and increase spectral resolution.  The new lenslet array increases the number of spaxels from 50$\times$50 to 73$\times$73, it increases the lenslet pitch from 360$\mu$m$\times$360$\mu$m to 500$\mu$m$\times$500$\mu$m, and it increases the length of each spectrum from 36 pixels to 80 pixels.  

The lenslet array is currently being fabricated by Jenoptik Optical Systems.  A previous version of the lenslet array (shown in Figure 4) had an error in the lenslet prescription.  The new lenslet array is expected to be delivered in August 2018.

\begin{figure}[htbp]
\begin{center}
  \hbox{
    \hspace{1.5in}
      \includegraphics[angle=0,width=0.5\linewidth]{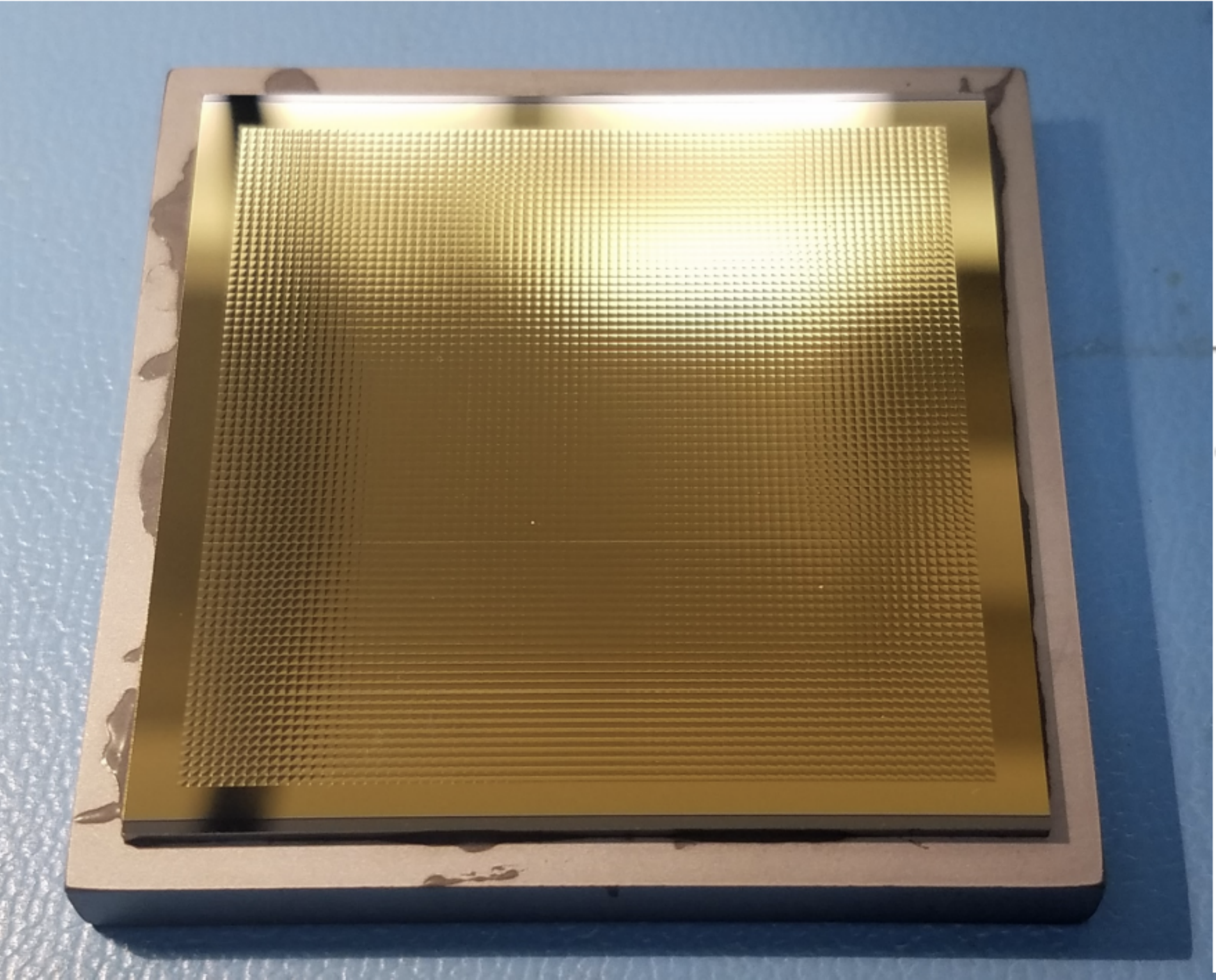}}
\end{center}
      \vspace{0.0in}
      \caption{--- The ALES upgrade involves a new lenslet array where each lenslet has a different optical prescription in order to pre-correct the downstream astigmatism from LMIRcam's single optic reimager.
	  } \label{fig:lenslet}
\end{figure}

\subsection{MAGNIFIERS}
The original implementation of ALES used a refractive magnifier.  In order to accommodate multiple wavelengths and platescales (see Table 1), we are switching to a set of reflective Keplerian magnifiers.  Each magnifier is a 1-inch diameter tube with gold coated primary and secondary mirrors.  Each of the LBT's two primary mirrors is imaged off-axis by the Large Binocular Telescope Interferometer.  Therefore, the secondary obscuration in the Keplerian magnifiers does not vignette either beam.  A cutout retrace of the generic magnifier design is shown in Figure 5.  The magnifiers, as fabricated, are shown in Figure 6.

\begin{deluxetable}{lcccccccccccc}
\tabletypesize{\small}
\tablecaption{Upgraded ALES Plate Scales and Fields of View}
\tablewidth{0pt}
\tablehead{
\colhead{Spaxel Plate Scale} &
\colhead{Field-of-View} &
\colhead{Nyquist Samples} &
\colhead{Nyquist Samples} &
\\
\colhead{} &
\colhead{} &
\colhead{(Single Aperture AO)} &
\colhead{(Dual Aperture Interferometry)} &
}
\startdata
6$\times$6 mas & 0.45$\times$0.45" & $>$0.5$\mu$m &  $>$1.4$\mu$m \\
12$\times$12 mas & 0.9$\times$0.9" & $>$1.0$\mu$m &  $>$2.8$\mu$m \\
25$\times$25 mas & 1.8$\times$1.8" & $>$2.0$\mu$m &   \\
50$\times$50 mas & 3.6$\times$3.6" & $>$4.0$\mu$m &   \\
\enddata
\label{specs}
\end{deluxetable}

\begin{figure}[htbp]
\begin{center}
  \hbox{
    \hspace{0.1in}
      \includegraphics[angle=0,width=0.9\linewidth]{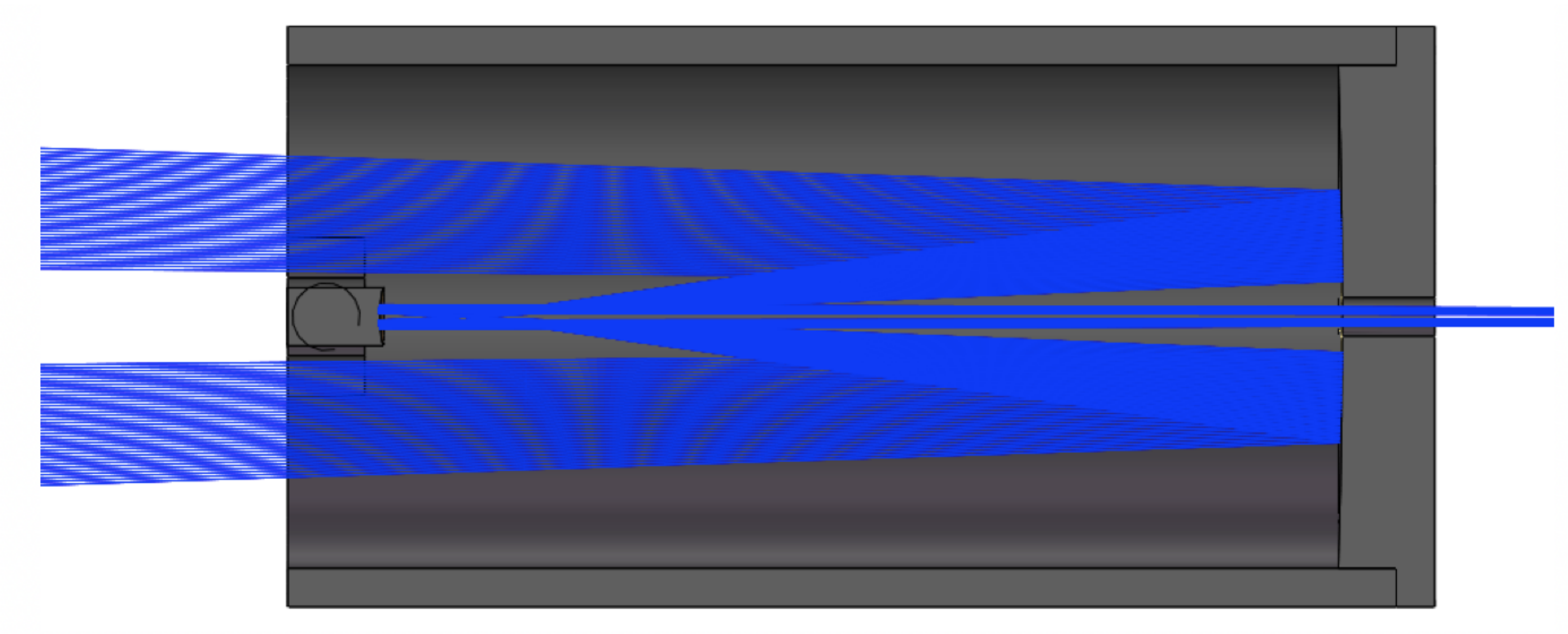}}
\end{center}
      \vspace{0.0in}
      \caption{--- Cutout SolidWorks schematic of magnifying optics with a Zemax ray-trace.  The beam enters on the left, reflects off of a primary and secondary, and then exits through a hole in the primary.  Because LBTI passes beams from the two LBT primaries off axis, there is no vignetting from the hole in the magnifier’s primary, or its secondary spider arms.
	  } \label{fig:mag1}
\end{figure}

\begin{figure}[htbp]
\begin{center}
  \hbox{
    \hspace{1.5in}
      \includegraphics[angle=0,width=0.5\linewidth]{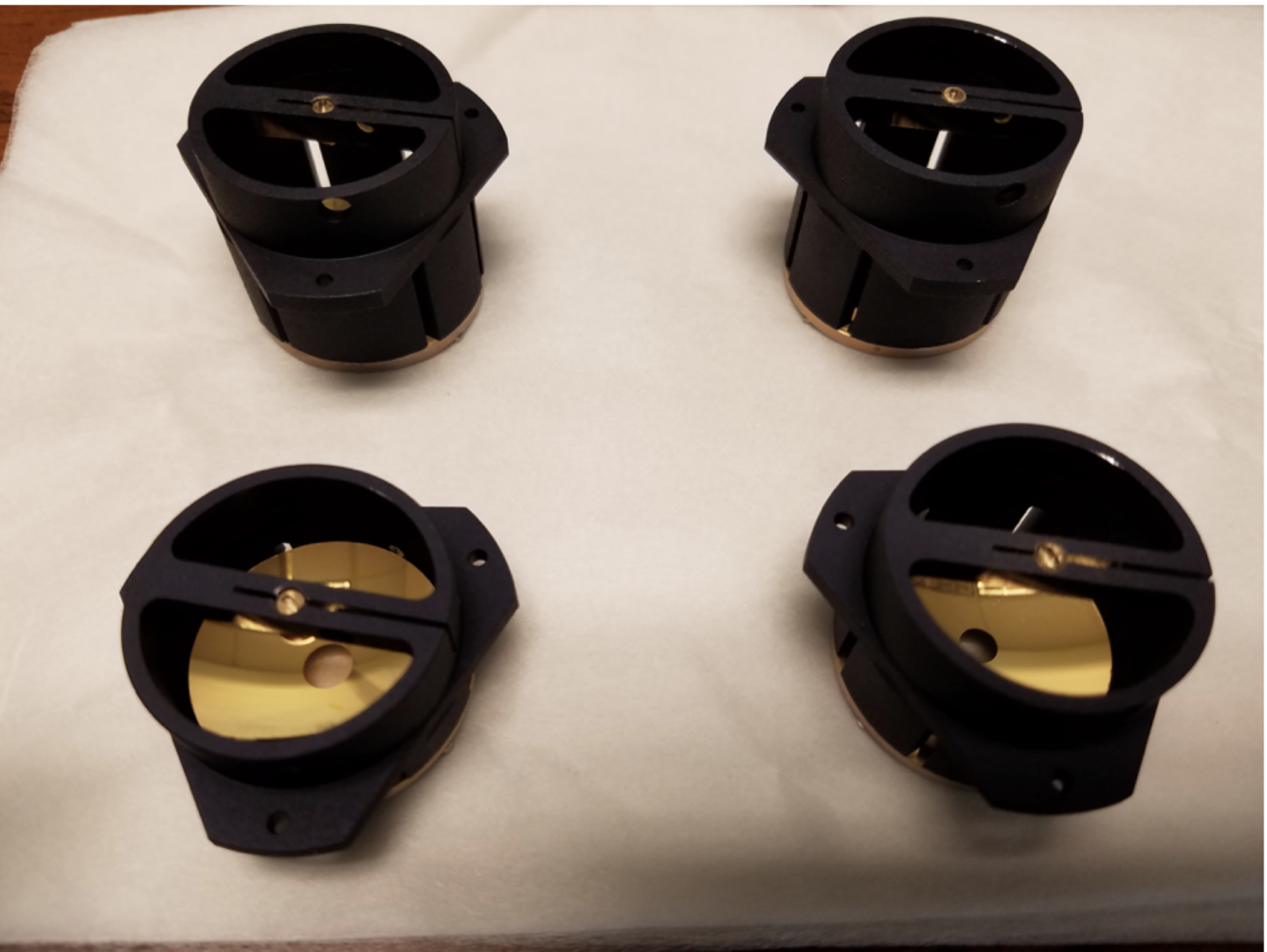}}
\end{center}
      \vspace{0.0in}
      \caption{--- The four ALES reflective magnifiers as built.
	  } \label{fig:mag2}
\end{figure}

\subsection{PRISM DISPERSERS}
The upgraded ALES will have 5 new prism dispersers (see Table 2), including a new version of the L-band disperser because the new ALES lenslet array enables longer spectra.  Each prism set is a sapphire - zinc selenide - sapphire direct vision prism, where the sapphire provides most of the dispersion and the zinc selenide diverts the beam back to a zero-deviation position.

The ALES prisms are installed at a pupil plane in a converging beam just before the LMIRcam detector. Prisms in converging beams create astigmatism and coma, which we mitigate by using minimum deviation (isosceles) prisms, and polishing a very slight cylinder into the zinc-selenide prism.

A schematic of a prism stack is shown in Figure 7.  One set of prisms has been delivered and will be tested with the new lenslet array.

\begin{deluxetable}{lcccccccccccc}
\tabletypesize{\small}
\tablecaption{Upgraded ALES Wavelengths and Spectral Resolutions}
\tablewidth{0pt}
\tablehead{
\colhead{Wavelength} &
\colhead{Spectral} &
\colhead{Mode} &
\\
\colhead{Range} &
\colhead{Resolution} &
\colhead{} &
}
\startdata
2.8-4.2$\mu$m & $\sim$40 & L-band \\
3.0-5.0$\mu$m & $\sim$20 & L/M-band \\
2.2-3.7$\mu$m & $\sim$40 & Ice-band \\
2.0-2.3$\mu$m & $\sim$150 & Ks / Br-$\gamma$ \\
3.1-3.5$\mu$m & $\sim$100 & PAH / CH$_{4}$ \\
\enddata
\label{specs2}
\end{deluxetable}

\begin{figure}[htbp]
\begin{center}
  \hbox{
    \hspace{1.5in}
      \includegraphics[angle=0,width=0.5\linewidth]{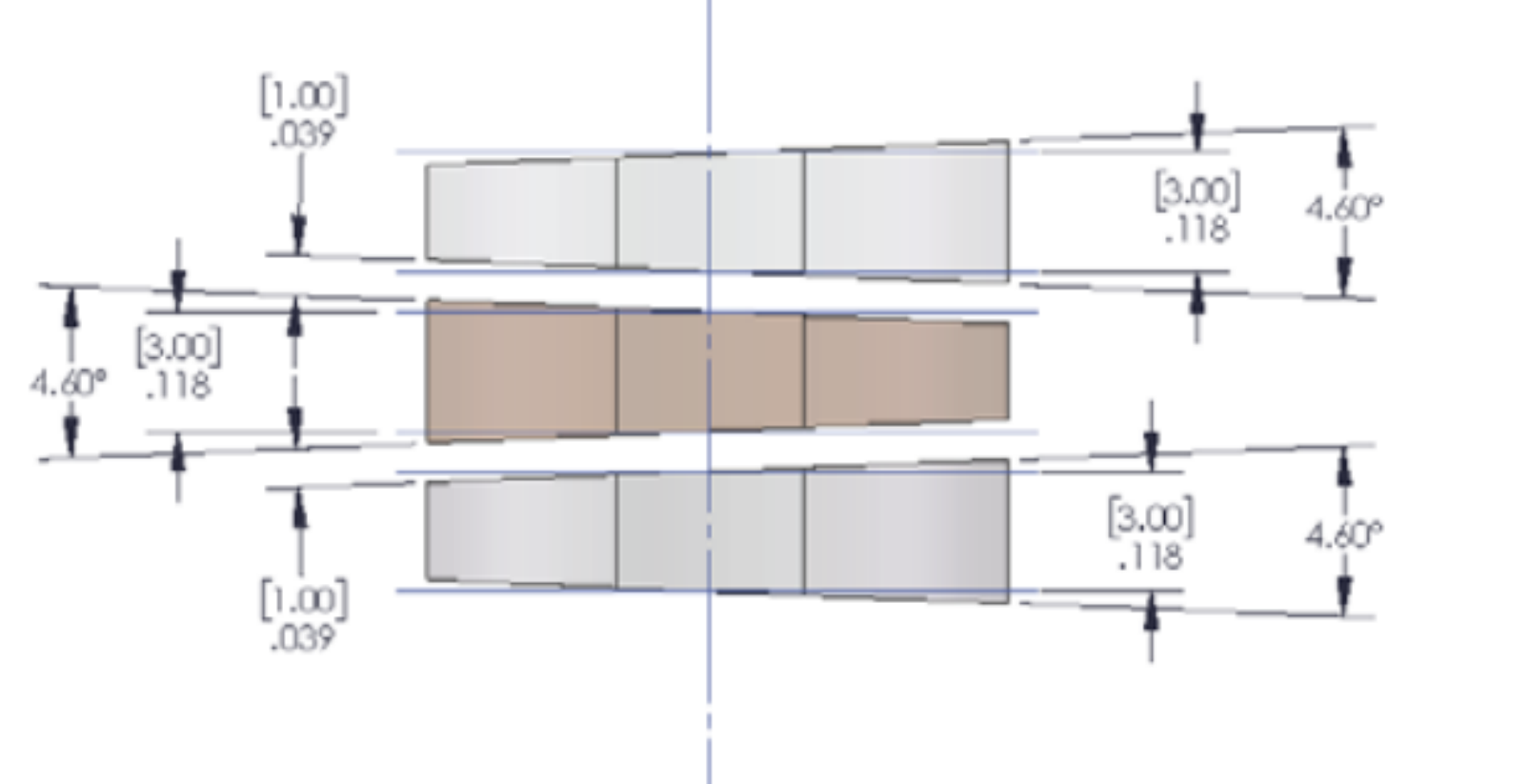}}
\end{center}
      \vspace{0.0in}
      \caption{--- Solidworks design of an ALES direct vision prisms, comprising a sapphire - zinc selenide - sapphire triplet.
	  } \label{fig:prism}
\end{figure}

\subsection{CALIBRATION}
Various aberrations in the ALES system make the point-spread-function of each lenslet different as a function of wavelength.  Additionally, because ALES is mounted in a filter wheel, the locations and aberrations of the spots could, in principle, change slightly.  The original ALES used narrowband filters to calibrate a wavelength solution and spot sizes for each individual lenslet.  We are currently building a monochromatic light injection unit to improve the ALES calibrations.  A lab setup of the system is shown in Figure 8.  The pipeline and calibration source are described in more detail in Briesemeister et al. (these proceedings)\cite{Briesemeister_2018SPIE}.

\begin{figure}[htbp]
\begin{center}
  \hbox{
    \hspace{0.1in}
      \includegraphics[angle=0,width=1.0\linewidth]{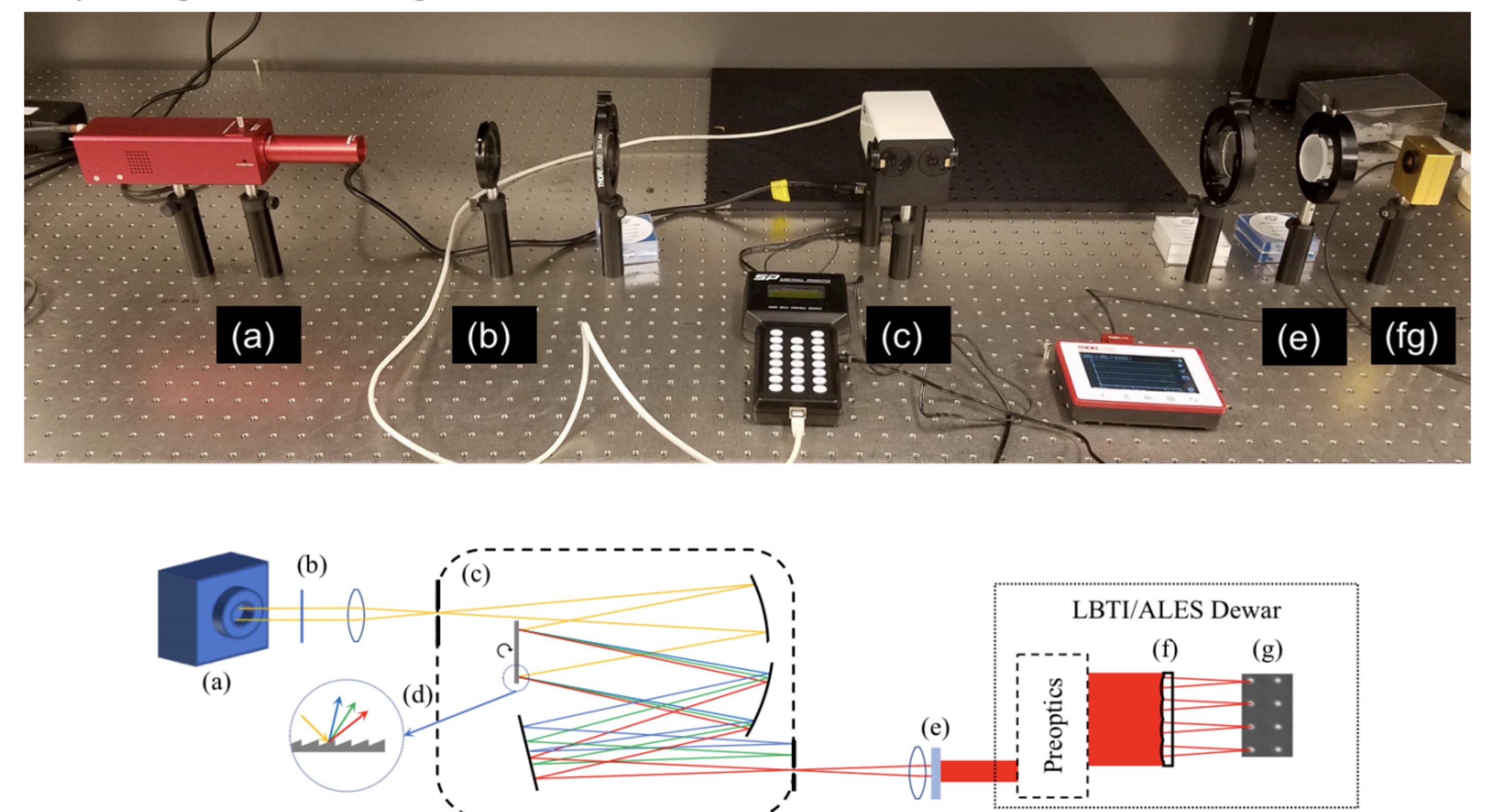}}
\end{center}
      \vspace{0.0in}
      \caption{--- A calibration system for ALES as implemented in the lab.  A thermal light source (a) passes a shutter (b), and is focused onto the slit of a monochrometer (c) with a grating selectable output wavelength (d), which illuminates a diffuser (e), which in turn illuminates the ALES lenslet array (f) and creates calibration spots on the LMIRcam detector (g).
	  } \label{fig:prism}
\end{figure}

\section*{ACKNOWLEDGMENTS}
This paper is based on work funded by NSF Grants 1608834 and 1614320.

\bibliography{database}
\bibliographystyle{spiebib}

\end{document}